\documentclass[twocolumn,amsmath,amssymb,floatfix,prl,superscriptaddress]{revtex4-1}
\usepackage{amsmath,amssymb,natbib,bm,graphicx,url,epsfig,psfrag}
\usepackage{amsmath,amssymb,natbib,bm,graphicx,url,epsfig,psfrag}
\usepackage{color}
\usepackage{ulem}

\begin{document}

\title{Noise on complex quantum Hall edges: Chiral anomaly and heat diffusion}

\author{Jinhong Park}
\affiliation{Department of Condensed Matter Physics, Weizmann Institute of Science, Rehovot 76100, Israel}
\author{Alexander D. Mirlin}
\affiliation{Institut f{\"u}r Nanotechnologie, Karlsruhe Institute of Technology, 76021 Karlsruhe, Germany}
\affiliation{Institut f{\"u}r Theorie der Kondenserten Materie, Karlsruhe Institute of Technology, 76128 Karlsruhe, Germany}
\affiliation{Petersburg Nuclear Physics Institute, 188300 St. Petersburg, Russia}
\author{Bernd Rosenow}
\affiliation{Institut f{\"u}r Theoretische Physik, Universit{\"a}t Leipzig, D-04103 Leipzig, Germany}
\author{Yuval Gefen} 
\affiliation{Department of Condensed Matter Physics, Weizmann Institute of Science, Rehovot 76100, Israel}
\date{\today}

\begin{abstract}

Electrical and thermal conductances of a quantum Hall bar reflect the topological structure of the incompressible bulk phase.  Here we show that noise of electrical current carried through the edge  evidences the interplay between these  two topological observables. Transport through a structured edge is modeled by a voltage-biased line junction made up of two counter-propagating modes associated with respective filling factors. Specifically, we focus on the edge of a $\nu=2/3$ fractional quantum Hall state.  Noise is generated at a point distinctly separated from the hot spot�� (where most of the Ohmic dissipation takes place) and reflects the competition between ballistically carried downstream current and diffusively carried heat (which can propagate also upstream). We propose specific setups where our predictions can be measured. 

\end{abstract}

\maketitle

Transport through quantum Hall edges has been intensively investigated both theoretically and experimentally over the last two decades.  One reason for this interest is that edge transport reflects the topological nature of the bulk through ��bulk-boundary correspondence��: interesting non-trivial facets of bulk topology, which are hardly directly accessible through bulk experiments, are manifested in edge physics. For example, measurements of shot noise of the electrical current on the edge passing through a quantum point contact prove the presence of fractionally charged quasiparticle excitations in the bulk~\cite{Heiblum}.

 Edge states of certain fractional filling factors turn out to have a complex structure: the edge hosts several chiral modes with different chiralities.
 The simplest example of such a system is the $\nu = 2/3$ fractional quantum Hall states. It was predicted that the edge of the $\nu = 2/3$ states consists of
 two counterpropagaing modes with filling factor discontinuities $\delta \nu = 1$ and $\delta \nu = -1/3$~\cite{MacDonald}. Kane, Fisher, and Polchinski showed that
 for a sufficiently strong interaction between the modes, intra-edge, inter-mode tunneling drives the system into an interaction- and disorder-dominated phase described by a downstream
 charge mode with $\delta \nu = 2/3$ and an upstream netural mode~\cite{KaneFisher}. 
These ideas have triggered experimental efforts, resulting in the discovery of edge neutral modes~\cite{Heiblum2}. In addition, it was recently noted that in the two terminal geometry, the electrical conductance varies from $G = 4e^2 / (3h)$ for a relatively short length between the contacts (coherent regime) to $G = 2e^2 / (3h) $ for a long length (incoherent regime)~\cite{Protopopov, Casey}, and this crossover of the conductance is experimentally measured~\cite{Yonatan}.
 
\begin{figure} [t]
\includegraphics[width=0.85\columnwidth]{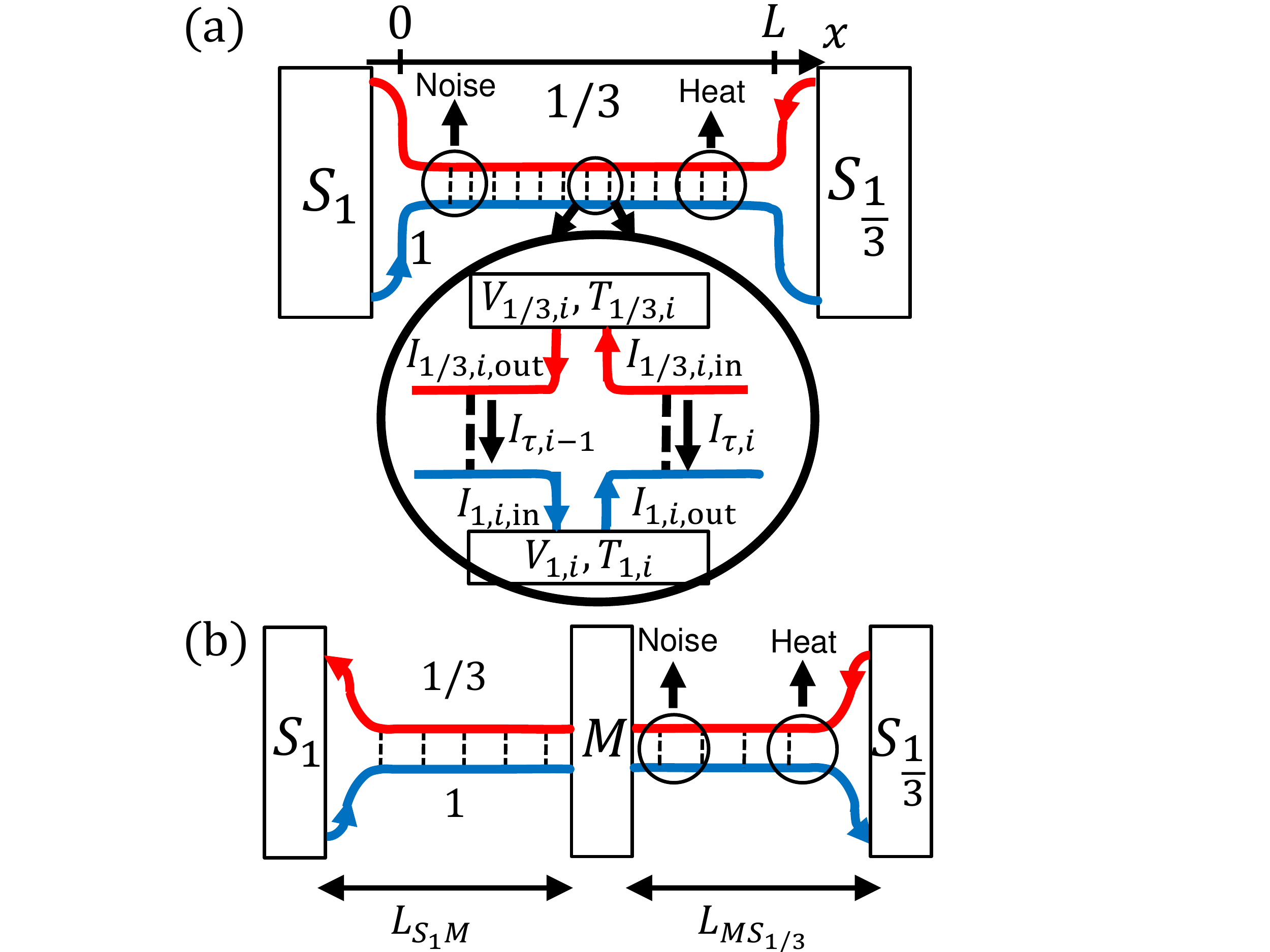} 
\caption{(color online) Experimental setups 
consisting of counter-propagating
$\delta\nu=1$ and $\delta\nu=-1/3$ edge modes, formed at the edge of the $\nu = 2/3$ fractional quantum Hall fluid. 
In each setup, locations of a hot spot (denoted as ``Heat'') and a noise-generating spot (denoted as ``Noise'') are marked.
(a) A noiseless voltage $V_0$ (voltage source) is applied at  contact $S_1$, and 
 current fluctuations are measured at the opposite contact $S_{1/3}$.  The dotted lines represent
tunneling between counter-propagating modes, the edge length is $L$.
Inset: Zoom-in on a section of the edge modes. Local virtual reservoirs on each mode are introduced to describe  local equilibration. The local voltage of reservoir $i$ of mode $\delta \nu = 1$ ($\delta \nu = -1/3$)
is determined in such a way that the incoming electrical current  $I_{1, i, \textrm{in}}$ ($I_{1/3, i, \textrm{in}}$) into reservoir $i$ is the same as the outgoing electrical current $I_{1, i, \textrm{out}}$ ($I_{1/3, i, \textrm{out}}$) out of  reservoir $i$. Equilibration between edge modes occurs via 
tunneling currents $I_{\tau,i}$.
(b) More realistic experimental setup. Similar to setup (a), a noiseless voltage $V_0$ is applied at $S_{1}$, but now 
 voltage fluctuations are measured at a middle contact $M$. The contact $M$ is floating with respect to voltage but at zero temperature thermally. 
 The length between $S_1$ ($S_{1/3}$) and $M$ is $L_{S_1M}$ ($L_{MS_{1/3}}$). }
\label{Setup}
\end{figure}

In addition to electrical transport, the thermal edge conductance reveals
the topology  of a bulk state~\cite{KaneFisher2}. In particular, recent measurements~\cite{Jezouin, Heiblumheat1, Heiblumheat2} of the thermal conductance of a variety of quantum Hall edges allow access to topological properties of the bulk states, not afforded by electrical conductance measurements. In the $\nu = 2/3$ state, the measured thermal conductance was $0.25$-$0.33 K_0 $~\cite{Heiblumheat1}, which appears to be a manifestation of the diffusive nature of heat transport~\cite{Protopopov, Casey} leading to zero heat conductance in the long edge limit.
Here, $K_0$ is $\pi^2 k_B^2  T  / (3h)$. 

In this Letter, we theoretically show that noise of the electrical current is characterized by the interplay between  electrical and  heat conductance.
Focusing on the case of the spin polarized $\nu = 2/3$ state, which gives rise to two counter propagating edge modes (corresponding to $\delta \nu = 1$ and $\delta \nu = -1/3$),
we represent the equilibration dynamics between these two modes employing the following picture. The edge is modeled by a line junction made up of
inter-mode tunneling bridges. In the incoherent regime, where the electrical conductance assumes the value of $G = 2e^2 / (3h)$~\cite{Protopopov, Casey}, we find that the noise is suppressed by a factor of $1/\sqrt{L}$, where $L$ is the distance between source and drain. This suppression of the noise originates from the geometrical separation
between the {\it hot spot} (where heat is generated) and the point where noise is generated. The dependence of the noise on $L$ is a manifestation of the interplay between chiral ballistic transport, universally determined by the chiral anomaly,  and diffusive propagation of heat, reflecting the bulk topological order. We propose an experimental scheme where this heat-noise separation can be detected. Such an analysis of the noise  would be useful to characterize distinct edge structures of different bulk topologies.


Our analysis of the noise generated in a line junction consisting of counter-propagating $\delta\nu=+1$ and $\delta\nu=-1/3$  
pertains to two experimental setups; one (Fig.~\ref{Setup}a) with two contacts ($S_1$ and $S_{1/3}$) and another (Fig.~\ref{Setup}b) with three contacts ($S_1$, $S_{1/3}$, and $M$).  The latter is more amenable to direct experimental detection. 
We  begin with a discussion of the the first setup, where  a noiseless voltage bias is applied at either $S_1$ or $S_{1/3}$, and the current fluctuations are measured at the opposite contact, $S_{1/3}$ or $S_1$, respectively.

Our main focus here is on the incoherent regime, where $L$ is larger than the equilibration length $\ell_{\textrm{eq}}$, over which one mode is equilibrated with the other~\cite{Protopopov}. Below we also discuss the short edge limit, where $L \ll \ell_{\textrm{eq}}$.
%
In the incoherent regime, the line junction is made up of segments, within each of which local chemical potentials 
$\mu(x) =eV_{1(1/3)}(x)$ and   local temperatures $T_{1(1/3)}(x)$ of either mode, $\delta \nu= 1 (-1/3)$ are defined. Such local equilibration is attained due to electron-electron interaction, but could be conveniently reached by introducing local virtual reservoirs between consecutive tunneling bridges~\cite{Casey}. $\mu(x)$ and $T(x)$ are then determined self-consistently, requiring that electrical current and energy flow into every given reservoir vanish~\cite{Buttiker1}.  Quantum interference  between different segments are neglected. We also ignore energy transport through the bulk~\cite{Gutman} or energy loss to an external bosonic bath (e.g. phonons)~\cite{Amit}.

\begin{figure} 
\includegraphics[width=\columnwidth]{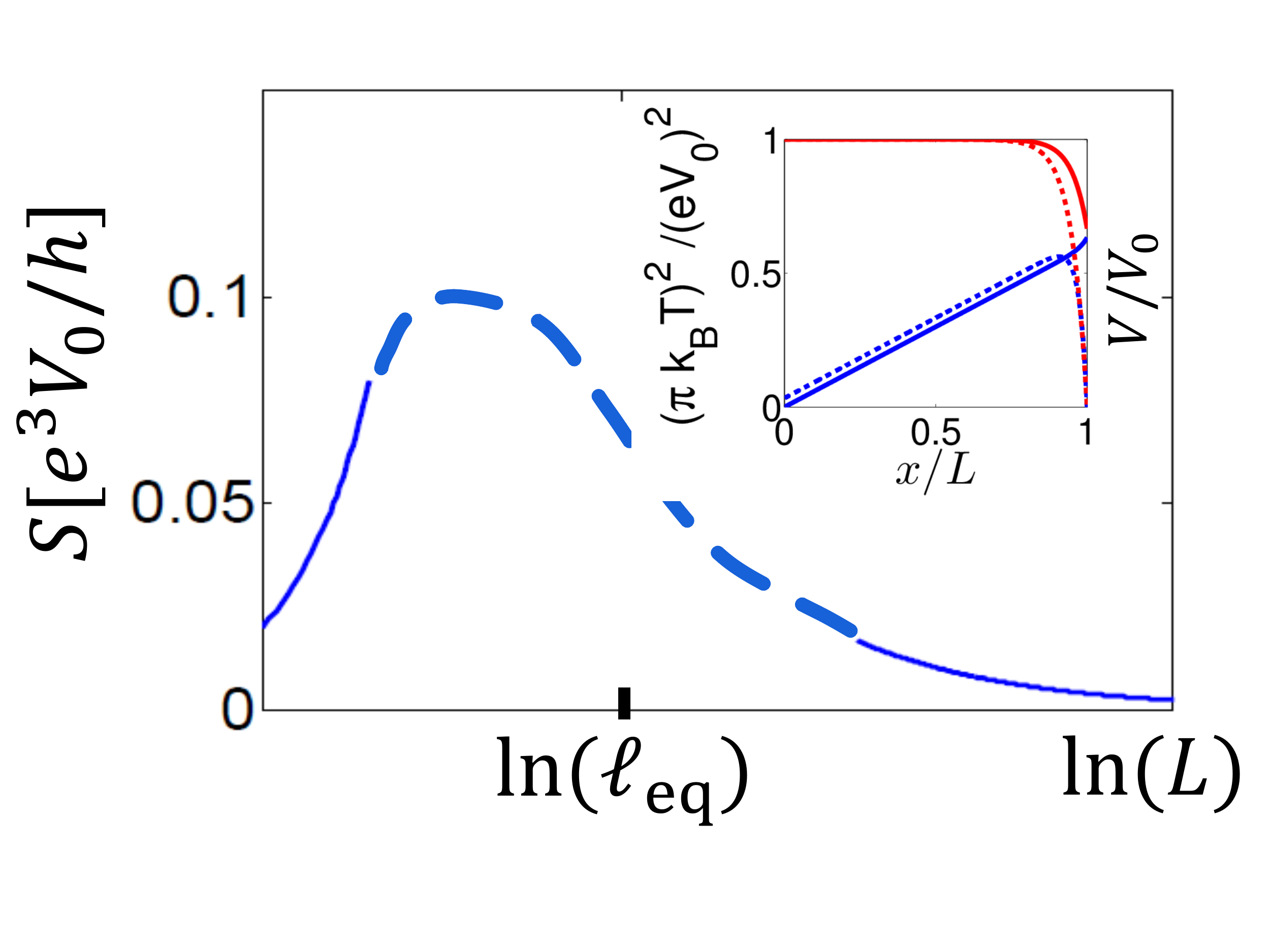} 
\caption{(color online) Schematic plot of intrinsic noise as a function of length $L$ (on a log-linear scale).
The solid curves correspond to the noise (Eqs.~\eqref{Noiseexpression} and \eqref{noisecoh}) calculated in 
the incoherent regime ($\ell_{\textrm{eq}} \ll L$) and
the coherent regime ($\ell_{\textrm{eq}} \gg L$).
With a dashed curve, we have schematically interpolated the solid curves to show crossover between two limiting regimes. 
Inset: Voltage and temperature profiles along the line junction with $L=10\ell_{\textrm{eq}} $. Red (blue) curves denote the voltage (respectively, temperature) profiles, with solid and dashed curves referring to the $\delta \nu = 1$ and $\delta \nu = - 1/3$ mode, respectively. Here we choose $\gamma = 9/5$, the value in the absence of interaction between the modes.
  }
\label{Noise_length2}
\end{figure}

Let us now bias the contact $S_1$ by a voltage $V_0$, leaving 
the contact $S_{1/3}$  grounded. The ambient temperatures
in the contacts are zero. As is apparent from the voltage profiles
(red curves in the inset of Fig.~2), the $\delta\nu = 1$ mode
 is equilibrated with the $\delta \nu=1/3$ mode within a distance $\ell_{\textrm{eq}}$ from $S_{1/3}$. Beyond this point  the voltages remain
constant, qualitatively similar  to a single chiral channel. The equilibration gives rise to dissipative heating near
$S_{1/3}$; a part of the generated heat is then transported diffusively to the contact
$S_1$; the diffusive nature of  heat transport is manifested through the linear growth of $T_1^2(x)$ and
$T_{1/3}^2(x)$ along (most of) the system (cf. the inset of Fig. 2.). The vicinity of $S_{1/3}$ where heating takes place is referred to as a {\it hot spot}.

\begin{figure} [t] 
\includegraphics[width=0.7\columnwidth]{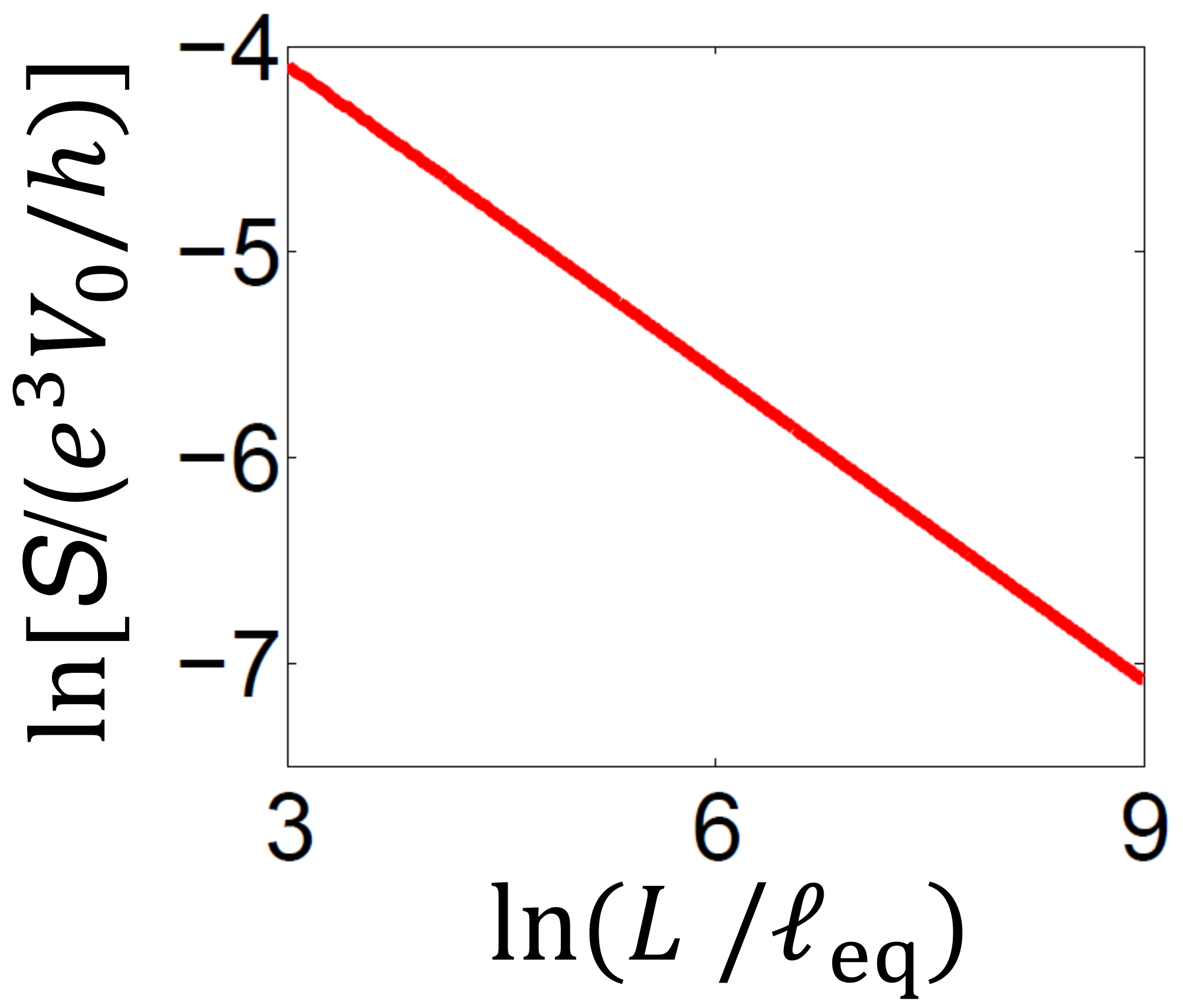} 
\caption{(color online) Noise (on a log-log scale) as a function of $L/\ell_{\textrm{eq}}$ ($L \gg \ell_{\textrm{eq}}$). This curve is
plotted as a result of the numerical calculation of the integral of Eq.~\eqref{Noiseexpression}.
The noise behaves as $\sqrt{\ell_{\textrm{eq}}/L}$. Here we use $\gamma = 9/5$.}
\label{Asymptotics}
\end{figure}

We study the non-equilibrium zero-frequency noise at $S_{1/3}$ in the setup depicted in Fig.~\ref{Setup}a via the zero frequency correlation function, $S \equiv \overline{(\delta I_{S_{1/3}})^2} $, of the current fluctuation ($\delta I_{S_{1/3}}$) at $S_{1/3}$~\cite{footnote}. For the noise $S$ we obtain 
\begin{align} \label{Noiseexpression}
S&\simeq \frac{8e^2}{h \ell_{\textrm{eq}}} \int_{0}^{L} dx \frac{e^{-4x/ \ell_{\textrm{eq}}}}{(3- e^{-2L/ \ell_{\textrm{eq}}})^2} \left [k_BT_{1}(x) + k_B T_{1/3}(x) \right ] \nonumber \\ 
& \simeq \frac{c e^2}{h} (eV_0) \sqrt{\frac{\ell_{\textrm{eq}}}{L}} \ , 
\end{align}
with $c =\left[ 2 \Gamma(3/2, 4/\gamma) e^{4/\gamma} +  \sqrt{\pi}\right] \sqrt{(2+\gamma)/6}   /(18 \pi )$.
Here, $\gamma$ is a dimensionless parameter which measures the deviation from the Wiedemann-Franz law for the tunneling heat current~\cite{Casey, KaneFisher3};
 in the absence of interaction between the modes, $\gamma = 9/5$ and $c \simeq 0.075$. $\Gamma (s, y)$ is the incomplete gamma function. 
 Employing the temperature profiles shown in the inset of Fig.~\ref{Noise_length2}, the noise is calculated and plotted as a function of edge length $L$ on a log-linear scale  in Fig.~\ref{Noise_length2} for both short $L \lesssim \ell_\textrm{eq}$ and and long $L \gg \ell_\textrm{eq}$ edges, and on a log-log scale in Fig. \ref{Asymptotics} for long edges. The $1/\sqrt{L}$ behavior differs from the $1/L$ behavior of the noise in a conventional metallic conductor~\cite{Nagaev, Kozub, Dejong, Supple}, where both electrical and heat currents are transported in a diffusive way.

 The behavior of the noise originates from the geometrical separation between the hot spot and noise-generating spots. As seen in 
 the integral over position $x$ of Eq.~\eqref{Noiseexpression}, the noise $S$ measured at $S_{1/3}$ is due to accumulation of the individual noise contributions generated in each tunneling bridge, but with weights depending on spatial positions. 
The tunneling bridges located within $\ell_{\textrm{eq}}$ from $S_{1}$  contribute most strongly to the noise $S$:
electrons, partitioned in tunneling bridges located within $\ell_{\textrm{eq}}$ from $S_1$, 
arrive at either $S_1$ or $S_{1/3}$, contributing to the noise $S$. On the other hand, 
all the electrons, partitioned in the tunneling bridges located beyond $\ell_{\textrm{eq}}$ from $S_1$, arrive at the same contact $S_{1/3}$, thus  not contributing to the zero frequency noise;
the electrons propagating on the $\delta \nu = -1/3$ mode after partitioning, are eventually backscattered at other tunneling bridges and arrive at $S_{1/3}$, 
along with the other electrons propagating on the $\delta \nu = 1$. 
This asymmetry regarding the location where the noise is generated becomes apparent in 
 the exponential factor of Eq.~\eqref{Noiseexpression}. Note that the location of the noise-generating spots are only determined by the direction of net chirality of the edge modes.
 Furthermore, the individual noise contributions generated in a tunneling bridge are dominantly governed by temperature of the modes at the bridge:
more heat arriving at the bridge generates more noise. Heat is generated near $S_{1/3}$, propagates in a diffusive way to $S_1$, and 
mostly contributes to the noise $S$ through the tunneling bridges in the vicinity of $S_1$. As $L$ increases, the noise-generating spots are further separated from the hot spot, a smaller amount of heat arrives at the noise-generating spots, and the noise measured at $S_{1/3}$ is suppressed. 

We now provide some technical details concerning our analysis.
We introduce $N$   virtual reservoirs on each edge mode to describe local equilibrium (cf. the inset of Fig.~\ref{Setup}a):
index $i$ of the reservoirs takes values from $1$ to $N$, moving from $S_1$ to $S_{1/3}$.
We assume that the attached reservoirs are ideal in such a way that there is no temporary charge accumulation in the reservoirs~\cite{Beenakker}; such a model is appropriate since a possible finite capacitance of the reservoirs would not modify the zero frequency noise.
Under this assumption, there is no net current flow between the edge modes 
and the reservoirs at any time $t$, and
\begin{equation} \label{currentconservationreservoir}
I_{1 (\frac{1}{3}), j,\textrm{out}} (t) = I_{1 (\frac{1}{3}), j, \textrm{in}}(t) 
\equiv I_{1 (\frac{1}{3}), j} \ .
\end{equation}
Here $I_{1 (1/3), j, \textrm{out}}$ is the outgoing current from reservoir $j$ and $I_{1 (1/3), j, \textrm{in}}$ is the incoming current to $j$ on the $\delta \nu = 1 ( -1/3)$ mode. As the ingoing and outgoing currents are equal to each other, we can drop the respective indices in the following. 
We can assign voltages to the edges via 
\begin{equation}
\label{outgoingcurrent}
 V_{1 , j} (t)      = \frac{h}{e^2}  I_{1, j} (t); \qquad
 V_{\frac{1}{3} , j} (t)   = \frac{3h}{e^2}     I_{\frac{1}{3}, j} (t), 
\end{equation}
whose differences will drive the tunnel currents $I_{\tau,i}$. 
Current conservation at each tunneling bridge implies 
\begin{align}  \label{currentconservationbridge}
  I_{1, j+1} (t) &= I_{1, j} (t)  + I_{\tau, j} (t), \nonumber \\ 
   I_{\frac{1}{3}, j+1} (t) &= I_{\frac{1}{3}, j} (t)  + I_{\tau,j} (t) \ .
\end{align}
In the following, we are interested in fluctuations of these currents, and for this reason we need to specify the statistical properties of the 
fluctuations in the $I_{\tau, j}$. It is useful to decompose the fluctuations according to 
\begin{align} \label{decomposition}
\delta I_{\tau ,j} = \delta I_{\tau, j}^{\rm tr}   + \delta I_{\tau ,j}^{\rm int},
\end{align}
where the transmitted fluctuations are determined by 
\begin{align} \label{transmittedCorrel}
\delta I_{\tau ,j}^{\rm tr} = g {e^2\over h} \left( \delta  V_{1/3,j+1} - \delta V_{1,j} \right).
\end{align}
Here, $g$ is the tunneling probability. We assume that $g$ is constant in different tunneling bridges and does not depend on the temperatures of the modes for simplicity.
The intrinsic contributions $\delta I_{\tau,j}^{\rm int}$ to  tunneling current fluctuations on the other hand are i.i.d. random variables with zero average
and variance (up to exponentially small contribution in the system size $L$)
\begin{align} \label{intrinsicCorrel}
\overline{\delta I_{\tau, j}^{\rm int}  \delta I_{\tau, j'}^{\rm int}} = 
 {2 e^2 \over h} g k_B (T_{1, j} + T_{\frac{1}{3}, j+1}) \delta_{j,j^\prime}.
\end{align}
We note that the shot noise contribution to $\overline{\delta I_{\tau, j}^{\rm int}  \delta I_{\tau, j'}^{\rm int}}$ is suppressed due to the fact that 
voltage differences between consecutive reservoirs are smaller than the respective temperatures
in the noise-generating spot (i.e., the region within $\ell_{\textrm{eq}}$ from $S_{1}$):
 voltage differences are exponentially suppressed as $\sim e^{-L/\ell_{\rm{eq}}}$, while the temperatures depend via a  power law $\sim \sqrt{\ell_{\rm{eq}}/L}$ on system size.
Using the above relations, Eqs.~\eqref{currentconservationreservoir}-\eqref{transmittedCorrel}, we can derive iterative equations for the fluctuations of the edge currents:
\begin{align}\label{iterative.eq}
\left(
\begin{array}{cc}  \delta I_{1/3,j+1} \\ \delta I_{1,j+1} \end{array} \right) = & {\bf \sf M}\left(
\begin{array}{cc}  \delta I_{1/3,j} \\ \delta I_{1,j} \end{array} \right) +  {\mathbf v}\, \delta I_{\tau,j}^{\rm int} \ .
\end{align}
Here we introduced the abbreviations
\begin{align}
{\bf \sf M} = & {1 \over 1 - 3g}
\left(\begin{array}{cc} 1 & - g \\ 3g & 1 - 4g \end{array}\right), \ 
{\mathbf v} =  {1 \over 1 - 3g} \left( \begin{array}{c} 1\\ 1 \end{array}\right).
\end{align}
Under the boundary conditions of $\delta I_{1, 0} = \delta I_{1/3, N+1} = 0$, 
Eq.~(\ref{iterative.eq}) can be solved as
\begin{align}\label{closedform.eq}
\left(
\begin{array}{cc}  0 \\ \delta I_{1,N+1} \end{array} \right) = & {\bf \sf M}^{N+1} \left(
\begin{array}{cc}  \delta I_{1/3,0} \\ 0 \end{array} \right) 
+ \sum_{j=0}^N {\bf \sf M}^{N-j} \,{\mathbf v} \,  \delta I_{\tau,j}^{\rm int} \ .
\end{align}
We finally obtain an explicit expression for the current fluctuation $\delta I_{S_{1/3}}$ at $S_{1/3}$ as
\begin{align}
\delta I_{S_{1/3}} =  \delta I_{1,N+1} =   - 2  \sum_{j = 0}^{N} \delta I_{\tau, j}^{\textrm{int}} \left (\frac{\eta^{N-j}}{1- 3 \eta^{N+1}}\right ),
\end{align}
where $\eta \equiv (1-g) / (1-3g)$.
Taking the continuum limit by sending $N \to \infty$, $g \to 0$ such that    $\ell_{\textrm{eq}} \equiv 2 L /(N \ln \eta)$ is finite, and using Eq.~\eqref{intrinsicCorrel}, 
the zero frequency noise $S \equiv \overline{(\delta I_{S_{1/3}})^2} $ measured at $S_{1/3}$ is given by Eq.~\eqref{Noiseexpression}. 

On the other hand, for  a short-length  line junction with $ L \ll \ell_{\textrm{eq}}  $, the conductance is $e^2 / h$ up to a correction by weak tunneling between the modes with total tunneling probability $P$ ($P \ll 1$)~\cite{Protopopov}. Under the assumption that the rare tunneling events are uncorrelated, the noise follows the usual expression 
of $S = 2 e^3 V_0 P / h$~\cite{Levitov, Feldman,Blanter}. The simple assumptions that $P \propto L$ and that $P$ does not depend on $V_0$ lead to 
\begin{align} \label{noisecoh}
S = \frac{2e^3 V_0}{h} \frac{L}{\ell_{\textrm{eq}}} \ .
\end{align}
The noise in the short and long junction limit is plotted in Fig.~\ref{Noise_length2} as a solid curve, an interpolation between these two limits is plotted as a dashed curve.

When the contact $S_{1/3}$ is biased by $V_0$ while the contact $S_1$ is grounded, the noise measured at $S_1$ has the same expression as Eq.~\eqref{Noiseexpression}.
 It comes from the fact that the noise measured at $S_1$ is the same as the noise measured at $S_{1/3}$ due to current conservation.
 Furthermore, the temperature profile in the case of biased $S_{1/3}$ is the exactly same as that in the case of biased $S_{1}$, and more 
 generally   
 the noise only depends on the voltage difference between the contacts.

We now turn our attention to the second setup depicted in Fig.~\ref{Setup}b, consisting again 
 of a noiseless voltage bias applied
at $S_{1}$ or $S_{1/3}$ as in the first setup, but we measure voltage fluctuations in the middle contact $M$, 
which is floating with respect to voltage but kept at zero temperature. 
We focus on the case of $\ell_{\textrm{eq}} \ll L_{MS_{1/3}}, L_{S_{1}M}$. 

Heat is generated only in the vicinity of $S_{1/3}$ as in the first setup. 
However, heat is not transported through  contact $M$ due to the assumption of keeping contact $M$ at zero temperature, and hence the temperatures of the modes between $S_1$ and $M$ are zero.
On the other hand, the temperature profiles of the modes between $M$ and $S_{1/3}$ are the same as the inset of Fig.~\ref{Noise_length2} with  $L$ replaced by $L_{M S_{1/3}}$. As a consequence, the fluctuations in the current $\delta I_{MS_{1/3}}$, generated in the line junction between $M$ and $S_{1/3}$,
 are the same as the fluctuations in $\delta I_{S_{1/3}}$  calculated previously, with $L$ replaced by $ L_{MS_{1/3}}$. Using the fact that the conductance of the incoherent $\nu=2/3$ edge is $(2/3) e^2/h$, the voltage fluctuations~\cite{Blanter, Supple} are  
\begin{align}
 \overline{(\delta V_M )^2}= \left ( \frac{3h}{2e^2}\right )^2  \overline{(\delta I_{MS_{1/3}})^2}\ .
\end{align}
Here, $\overline{(\delta I_{MS_{1/3}})^2}$ is given by  Eq.~\eqref{Noiseexpression} with $L$ replaced by $L_{MS_{1/3}}$.
We note that the voltage fluctuations in the middle contact reflect the intrinsic noise generated in the line junction between $M$ and $S_{1/3}$. In the experiment~\cite{yaron} with the same contact configuration as described here, finite voltage fluctuations in the middle contact with fixed $L_{MS_{1/3}} \simeq 25\mu m$
 are reported, but the magnitude of voltage fluctuations as a function of $L_{MS_{1/3}}$ needs to be measured in a future experiment for verifying our predictions.

 In summary, we study intrinsic noise generated in a line junction consisting of two counter-propagating modes, a $\delta \nu = 1$ mode and a $\delta \nu = -1/3$ mode. 
 We find a novel mechanism to generate  noise in the incoherent regime for a specific setup (cf. Fig.~\ref{Setup}a); 
 heat is generated near the contact out of which the $\delta \nu = -1/3$ mode emerges, is then transported in a diffusive way  through the line junction to the other contact, 
 and finally generates  noise in the vicinity of the latter contact. This is a result of the interplay between two topological properties: electrical and thermal transport coefficients.
 In a more realistic setup (cf. Fig.~\ref{Setup}b), the noise is reflected in the voltage fluctuations of a middle contact. 
 This analysis can be generalized to the other hole-conjugate quantum Hall states with the filling factor $1/2 < \nu <1$.
 Experimental  platforms with line junction systems consisting of counterpropagating modes (one of which is a fractional edge) were
experimentally realized at the interface between quantum Hall regions with different bulk filling factors~\cite{Anna, Ronen} as well as at the edge of  hole-conjugate quantum Hall states.

\begin{acknowledgements}

We thank  Ivan Protopopov, Kyrylo Snizhko, Eugene Sukhorukov, Yonatan Cohen, Wenmin Yang, and Moty Heiblum for helpful discussions. 
ADM and YG acknowledge support by DFG grant No. MI 658/10-1. 
BR and YG acknowledge support by DFG grant No. RO 2247/8-1. 
YG acknowledges support from CRC 183 of the DFG, ISF Grant No. 1349/14, and Italia-Israel project QUANTRA.

\end{acknowledgements}

\onecolumngrid

\setcounter{equation}{0}
\setcounter{figure}{0}
\setcounter{table}{0}
\setcounter{page}{1}
\renewcommand{\theequation}{S\arabic{equation}}
\renewcommand{\thefigure}{S\arabic{figure}}
\renewcommand{\bibnumfmt}[1]{[S#1]}
\renewcommand{\citenumfont}[1]{S#1}

\section{supplemental material}

The orginaization of this Supplemental Material is as follows. In Section A, we calculate the voltage and temperature profiles on a line junction consisting of $\delta \nu = 1$ and $\delta \nu = - \nu = -1/(2m+1)$ with positive integer $m$. The $\delta \nu = -\nu$ mode generalizes the $\delta \nu = -1/3$ mode considered in the main text. In Section B, we derive noises
(Eqs.~(1) and (13) in the main text) generated in the line junction with a variety of boundary conditions in contacts. In Section C, we consider a line junction consisting of two counter-propagating modes $\delta \nu = 1$ and $\delta \nu = -1$ and derive the noise generated in the line junction, reproducing the well known result found in some literatures~\cite{Nagaev, Kozub, Dejong}. We contrast the noise with that of the line junction considered in the main text.  Finally, 
we address a setup with five contacts in Sec.~D.

\section{a. voltage and temperature profiles on a line junction}

\begin{figure} [h]
\includegraphics[width=0.7\columnwidth]{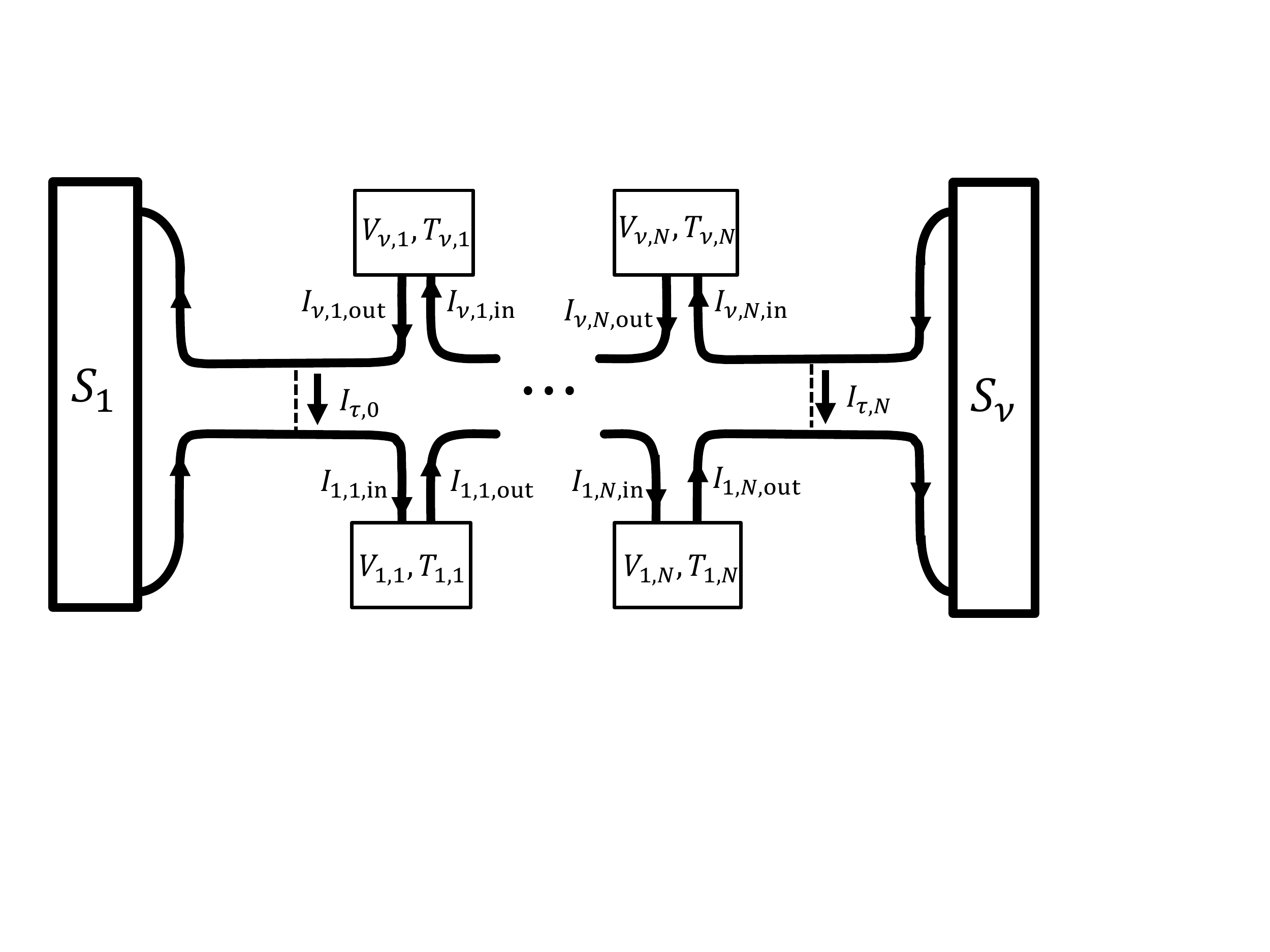} 
\caption{A line junction consisting of counter-propagating
$\delta\nu=1$ and $\delta\nu=-\nu$ edge modes. The length of the line junction is $L$.
A noiseless voltage $V_0$ is applied at contact $S_1$ or $S_{1/3}$ and current (or current fluctuation) is measured at the opposite contact $S_{\nu}$ or $S_{1}$, respectively. We attach artificial probe reservoirs to each edge mode for describing local equilibrium of the mode. The voltage (temperature) of each reservoir is determined in such a way that the incoming electrical (energy) current into the reservoir is the same as the outgoing electrical (energy) current our of the reservoir. Equilibration between the edge modes occurs via tunneling currents $I_{\tau, j}$ denoted by dashed lines.
} \label{LinejunctionMultiple} 
\end{figure}


In this section, we compute the voltage and temperature profiles of two counter-propagating 
edge modes, the $\delta \nu = 1$ mode and the $\delta \nu =- \nu = -1/(2m+1)$ mode (for positive integer $m$) along a line junction, depicted in Fig.~\ref{LinejunctionMultiple}.
Some results in this section are overlapped with Ref.~\cite{Casey}.
The $\delta \nu = -\nu$ mode generalizes the $\delta \nu = -1/3$ mode considered in the main text. 
We attach $2N$ virtual reservoirs ($N$ on each edge mode) for conveniently attaining local equilibration of the modes; The local voltage $V_{1, j}$ ($V_{\nu, j}$) and temperature $T_{1, j}$ ($T_{\nu, j}$) of reservoir $j$ on the $\delta \nu = 1$ ($\delta \nu = -\nu $) mode are self-consistently determined requiring that electrical current and energy
flow into reservoir $j$ vanish. Following the same procedure as Eqs.~(2)-(6) in the main text, 
we set up equations for reservoir $j$ ($1\leq j \leq N$) on the $1$ and $\nu$ modes as 
\begin{align} \label{currentconservation1}
 I_{1, j+1} (t) &=  I_{1 ,j} (t)  + \left [ g \left (I_{\nu, j+1}(t) / \nu -  I_{1, j} (t) \right ) +  I_{\tau, j}^{\rm{int}} (t) \right ] , \\ 
\label{currentconservation1/3}
I_{\nu, j+1} (t) &= I_{\nu, j} (t) +  \left [ g \left (I_{\nu, j+1}(t) / \nu -  I_{1, j} (t) \right ) +  I_{\tau, j}^{\rm{int}} (t) \right ] .  
\end{align}
Employing time averages of $I_{1, j} = e^2 \overline{V_{1,j}}/h$, $I_{\nu, j}=\nu e^2 \overline{V_{\nu,j}}/h$, and $\overline{I_{\tau, j}^{\rm{int}}} = 0$, we obtain 
\begin{align} \label{voltageEq}
  \frac{\nu e^2}{h} \overline{V_{\nu, j}} =  (\nu - g) \frac{e^2}{h} \overline{V_{\nu, j+1}}+g\frac{e^2}{h} \overline{V_{1, j}},
\nonumber \\ 
  \frac{e^2}{h} \overline{V_{1, j}} = (1 - g) \frac{e^2}{h} \overline{V_{1, j-1}}+g\frac{e^2}{h} \overline{V_{\nu, j}}. 
\end{align}
  Here $\overline{O(t)} \equiv \int_{0}^{\tau} O(t') dt' / \tau$ is the average value of $O$ in time. 

Let us consider the case that the contact $S_{1}$ is biased by a voltage $V_0$, leaving the contact $S_{\nu}$ grounded. Under the corresponding
boundary conditions of $\overline{V_{1, 0}} = V_0$ and  $\overline{V_{\nu, N+1}} = 0$, Eq.~\eqref{voltageEq} can be solved as 
\begin{align} \label{chemicalpotentialDiscrete}
\overline{V_{1, j}} &= V_0 \frac{ \nu \eta^j- \gamma^{N+1}}{\nu - \eta^{N+1}},\,\,\,\,\,\,\, \overline{V_{\nu, j}}= V_0 \frac{\eta^j- \eta^{N+1}}{\nu - \eta^{N+1} },
\end{align}
where $\eta = (1-g) /(1 - g/ \nu)$. Taking the continuum limit by sending $N \rightarrow \infty $ and
defining the equilibration length $\ell_{\textrm{eq}}$ (over which one mode is  equilibrated with the other) as 
$\ell_{\textrm{eq}} = 2 L/(N \ln \eta) = 2 L / [N \ln [1+ g(1-\nu)/ (\nu - g)]] \simeq 2  \nu L / [N g( 1- \nu)]$, the voltage profiles of the modes
are written as 
\begin{equation}  \label{VoltageProfile_S1}
V_{1} (x) =V_0 \frac{1 - \nu e^{2(x - L) /\ell_{\textrm{eq}}}}{1-  \nu e^{- 2L / \ell_{\textrm{eq}}}},\,\,\,\,\,\,\,\,\,V_ \nu (x) =V_0 \frac{1 - e^{2(x - L) /\ell_{\textrm{eq}}}}{1-  \nu e^{- 2L /\ell_{\textrm{eq}}}}.
\end{equation}
These voltage profiles of the modes for $\nu = 1/3$ are drawn in the inset of Fig.~2. 
The net current $I_{S_\nu}$ measured at $S_{\nu}$ is 
\begin{equation}
I_{S_{\nu}} = \frac{e^2 \overline{V_{1, N}}}{h} (1- g) = \frac{e^2 V_0}{h} (1-\nu) \frac{\eta^{N+1}}{\eta^{N+1} - \nu} \xrightarrow{N \rightarrow \infty} \frac{e^2 V_0}{h}  \frac{ 1-\nu}{1-\nu e^{-2L / \ell_{\textrm{eq}}}}.
\end{equation}
When $S_{\nu}$ is biased by $V_0$ while $S_{1}$ is grounded, the voltage profiles are  
\begin{equation} \label{VoltageProfile_Snu}
V_{1} (x) =\nu V_0 \frac{ e^{2(x - L) /\ell_{\textrm{eq}}} - e^{-2L / \ell_{\textrm{eq}}}}{1-  \nu e^{- 2L / \ell_{\textrm{eq}}}},\,\,\,\,\,\,\,\,\,V_ \nu (x) =V_0 \frac{ e^{2(x - L) /\ell_{eq}} - \nu e^{-2L / \ell_{\textrm{eq}}}}{1-  \nu e^{- 2L /\ell_{\textrm{eq}}}}.
\end{equation}

Let us move on to the temperature profiles of the modes.
The temperature profiles are determined such that the energy current flowing into each reservoir is the same as the energy current flowing out of the reservoir. Together with energy conservation in tunneling bridges, it leads to 
\begin{align}
\frac{e^2  \overline{V_{1, j}}^2 }{2h}  + \frac{\pi^2 k_B^2 T_{1, j}^2}{6 h}  &= \Big(\frac{e^2  \overline{V_{1, j-1}}^2 }{2h} (1 -  g) + \frac{\pi^2 k_B^2 T_{1, j-1}^2}{6 h} (1- \gamma g) + 
\frac{e^2 \overline{V_{\nu, j}}^2}{2h}g +\frac{\pi^2 k_B^2 T_{\nu, j}^2}{6 h}\gamma g
\Big), \nonumber \\
\frac{\nu e^2  \overline{V_{\nu, j}}^2 }{2h}  + \frac{\pi^2 k_B^2 T_{\nu, j}^2}{6 h}  &= \Big(\frac{e^2  \overline{V_{\nu, j+1}}^2 }{2h} (\nu - g) + \frac{\pi^2 k_B^2 T_{\nu, j+1}^2}{6 h} (1-\gamma g) + 
\frac{e^2 \overline{V_{1, j}}^2}{2h} g +\frac{\pi^2 k_B^2 T_{1, j}^2}{6 h} \gamma g
\Big).
\end{align}
Here $\gamma$ is a phenomenological parameter to measure the deviation from the Wiedemann-Franz Law;
for non-interacting fermions, $\gamma = 1$ and the Wiedemann-Franz law holds. In the absence of interaction between the modes, $\gamma$
 becomes $3/(2\nu+1)$; for the derivation, see Ref.~\cite{Casey}. 
Employing the boundary conditions of $T_{1, 0} = T_{\nu, N+ 1} = 0$, 
the temperature profiles of the modes along the line junction can be obtained as 
\begin{align} \label{tempprofile}
k_B^2 T_{1} ^2(x) &=  \frac{3  \nu e^2 V_0^2  }{2 \pi^2}\frac{1}{(1-  \nu e^{- 2L /\ell_{\textrm{eq}}})^2}\bigg [\frac{2 \nu \gamma x}{2 \nu \gamma L +(1 - \nu)\ell_{\textrm{eq}}} [(1 - \nu + \gamma \nu)+ (1-\nu - \gamma \nu) e^{-4L /\ell_{\textrm{eq}}}]
\nonumber \\ & +  (1-\nu-\gamma \nu)  (e^{4(x- L) / \ell_{\textrm{eq}}}- e^{-4 L / \ell_{\textrm{eq}}})\bigg ],\nonumber \\
k_B^2 T_{\nu} ^2(x) &= -\frac{3  \nu e^2 V_0^2  }{2 \pi^2}\frac{1}{(1-  \nu e^{- 2L /\ell_{\textrm{eq}}})^2}\bigg [\frac{2 \nu \gamma (L-x)}{2 \nu \gamma L +(1 - \nu)\ell_{\textrm{eq}}} [(1 - \nu + \gamma \nu)+ (1-\nu - \gamma \nu) e^{-4L /\ell_{\textrm{eq}}}]
\nonumber \\ & - ( 1  - \nu +\gamma \nu) (1-  e^{4(x- L) /\ell_{\textrm{eq}}})\bigg ].
\end{align}
Note that the temperature profiles in the case of biased $S_1$ are the exactly same as those in the case of
$S_{\nu}$; for both cases, heat is generated only in the vicinity of $S_{\nu}$ and a part of generated heat is transported diffusively to $S_1$.
The temperature profiles of the modes with $\nu  = 1/3$ and $\gamma = 3/(2\nu+1) = 9/5$ are drawn in the inset of Fig.~2.

\section{b. Excess noise at multiple bridges}
In this section, we derive Eqs.~(1) and (13) (in the main text) with more details and check the Johnson-Nyquist noise in the equilibrium case (of contacts with the same ambient temperature).
We compute the current fluctuation measured at drain in several cases of contacts: (i) 
a noiseless voltage is applied at contacts (under zero temperature), (ii) finite temperature is kept for both contacts, and (iii) fluctuating voltages are applied at both contacts (under zero temperature).
Here, we focus on the incoherent regime that the equilibration length $\ell_{\textrm{eq}}$ is much smaller than $L$.

We solve Eqs.~\eqref{currentconservation1} and \eqref{currentconservation1/3} under general boundary conditions without any concrete conditions of $\delta I_{1, 0} (t)$ and $\delta I_{\nu, N+1} (t) $. 
The fluctuation $\delta I_{S_{\nu}} (t) = I_{S_{\nu}} (t) - \overline{I_{S_{\nu}}(t)}$ of the net current measured at $S_{\nu}$ is calculated as 
\begin{align} \label{currentfluc}
\delta I_{S_{\nu}} (t) &= \delta I_{1, N+1} (t)  - \delta I_{\nu, N+1} (t)
=  (1- \nu)  \sum_{j = 0}^{N}\left [ \frac{\eta^{N-j} \delta I_{\tau, j}^{\textrm{int}} -  \delta I_{\nu, N+1} + \eta^{N+1} \delta I_{1, 0}}{ \eta^{N+1} - \nu }\right].
\end{align}
We put $\delta$ in order to represent deviation from the average value of $O$ ($\delta O(t) = O(t) - \overline{O(t)}$).

Let us consider the case (i) addressed in the main text:  i.e., $\delta I_{1, 0} = \delta I_{\nu, N+1} = 0$.
The noise at $S_{\nu}$ is written as 
\begin{align}
\overline{(\delta I_{S_{\nu}})^2} \simeq & \frac{(\nu- 1)^2 }{(\nu - \eta^{N+1})^2} \sum_{j=0}^{N} \overline{\left (\delta I_{\tau, j}^{\rm{int}}\right )^2}\eta^{2(N - j)},
\end{align}
and then in the continuum limit ($N \rightarrow \infty$), it becomes
\begin{align}  \label{expressionofnoise}
\overline{(\delta I_{S_{\nu}})^2}  &= \lim_{N\rightarrow \infty} \sum_{j=0}^{N} \Delta x_{j}\frac{N+1}{L}\frac{(\nu- 1)^2  \eta^{2(N  - j)} }{(\nu - \eta^{N+1})^2} 
  \overline{\left (\delta I_{\tau, j}^{\rm{int}}\right )^2}
\nonumber \\ &= \frac{4e^2 }{h} \frac{\nu (1-\nu)}{\ell_{\textrm{eq}}} \int_{0}^{L} dx \frac{e^{-4x/\ell_{\textrm{eq}}}}{(1- \nu e^{-2L/\ell_{\textrm{eq}}})^2} [k_BT_{1}(x) + k_B T_{\nu}(x)]. 
\end{align}
Here we have used 
\begin{align} \label{noisetunnelingbridges}
\overline{\delta I_{\tau, j}^{\rm{int}}\delta I_{\tau, j'}^{\rm{int}}} &=  \frac{2 e^2}{{h}} g (\overline{V_{1, j-1}} -\overline{ V_{\nu, j}})  \coth \Big ( \frac{e(\overline{V_{1, j-1}} - \overline{V_{\nu, j}}) }{ k_B (T_{1, j-1} + T_{\nu, j})} \Big ) \delta_{jj'}.
\end{align}
The fact that the voltage difference between consecutive reservoirs is much smaller than the respective temperatures ($(\overline{V_{1, j-1}} - \overline{V_{\nu, j}}) \gg (T_{1, j-1} + T_{\nu, j})/2$)
in the noise-generating spot (i.e., the region within $\ell_{\rm{eq}}$  from $S_1$)
simplifies Eq.~\eqref{noisetunnelingbridges} as $\overline{\delta I_{\tau, j}^{\rm{int}}\delta I_{\tau, j'}^{\rm{int}}} \simeq  2 e^2 g \left [ k_B (T_{1, j-1} + T_{\nu, j}) \right ] \delta_ {j, j'} / h$.
When $\nu = 1/3$, Eq.~\eqref{noisetunnelingbridges} becomes Eq.~(1) in the main text. 

Note that the noise generated in the line junction only depends on the temperature profiles of the modes, dictated by the absolute value of the voltage difference between the contacts (see Eq.~\eqref{tempprofile} and the text below Eq.~\eqref{tempprofile}); the noise is symmetric with respect to flipping the sign of the voltage in one of the contacts. Furthermore, the noise does not depend on which contacts are biased or grounded because
 the temperature profiles in the case of biased $S_1$ (and grounded $S_{\nu}$) are the exactly same as those in the case of
$S_{\nu}$ (and grounded $S_{1}$).

Now, we move on to the case (ii) of contacts with ambient temperature $T$. Then, the temperature profiles along the line junction are constant with $T_1 (x) = T_{\nu} (x) = T$. 
Using the relations of $\overline{(\delta I_{1, 0})^2} = 2 e^2 k_B T /h$,
$\overline{(\delta I_{\nu, N+1})^2} = 2 e^2 \nu k_B T /h$, $\overline{\delta I_{1, 0} \delta I_{\tau, j}^{\rm{int}}} = 2 e^2  k_B T  g \delta_{j, 0}/h$ and 
$\overline{\delta I_{\nu, N+1}\delta I_{\tau, j}^{\rm{int}}} = 2 e^2 k_B T g \delta_{j, N}/h$,
the noise at $S_{\nu}$ is computed as 
\begin{align} 
\overline{(\delta I_{S_{\nu}})^2}  &\simeq \frac{4e^2 (1 - \nu)}{h  \ell_{\textrm{eq}} \nu} \int_{0}^{L} dx \frac{e^{-4x/\ell_{\textrm{eq}}}}{(1/\nu- e^{-2L/ \ell_{\textrm{eq}}})^2} [k_BT_{1}(x) + k_B T_{\nu}(x)]+ \frac{2e^2}{h} (1 - \nu)^2 k_B T,
\nonumber \\ & \simeq \frac{2e^2}{h}(1 - \nu) k_B T = 2 G k_B T.
\end{align}
Here $G$ is the conductance from $S_1$ to $S_{1/3}$. This noise coincides with the Johnson-Nyquist noise.

Finally, we consider the case (iii) that the voltages of contacts are fluctuating in time: $\delta V_{1, 0} (t)$ and $\delta V_{\nu, N+1} (t) $ are not zero.
Then the current fluctuations at $S_{1}$ and $S_{\nu}$ are given as
\begin{align}
\Delta I_{S_{1}} &= \Delta I_{S_{\nu}} = \delta I_{S_{\nu}} + \frac{e^2}{h} \frac{ 1- \nu}{1 - \nu e^{- 2L /\ell_\textrm{eq}}} \left [ \delta V_{1} ( x= 0, t) - \nu  e^{- 2L /\ell_\textrm{eq}} \delta V_{\nu} (x = L, t)
 \right ], \nonumber \\ 
 & \xrightarrow{\ell_{\textrm{eq}} \ll L}  \delta I_{S_{\nu}} + \frac{e^2}{h} \left( 1- \nu \right ) \delta V_{1} (x= 0, t).
\end{align}
We apply this formula to the setup with three contacts displayed in Fig.~1(b) in the main text. 
Heat is generated only in the vicinity of $S_{1/3}$. 
Under the assumption that the heat is not transported through the contact $M$, temperatures of the modes between $S_1$ and $M$ are zero.
On the other hand, the temperature profiles of the modes between $M$ and $S_{1/3}$ are the same as the inset of Fig.~2 in the main text with  $L$  replaced by $L_{M S_{1/3}}$.
There are two contributions to the fluctuating component $\Delta I_{MS_{1/3}}$ ($\Delta I_{S_1M}$) of the current propagating along the line junction
 between $M$ ($S_1$) and $S_{1/3}$ ($M$):
the intrinsic current fluctuation $\delta I_{MS_{1/3}}$ ($\delta I_{S_1M}$) generated in the line junction between $M$ ($S_{1}$) and $S_{1/3}$ ($M$), and the current fluctuation due to voltage fluctuation of the contact $M$ ($S_{1}$),
\begin{align}
\Delta I_{MS_{1/3}} &= \delta I_{MS_{1/3}} + \frac{2e^2}{3h} \delta V_{M}, 
\nonumber \\ 
\Delta I_{S_1M} & = \delta I_{S_1M} + \frac{2e^2}{3h} \delta V_{S_1}.
\end{align}
The voltage fluctuation $\delta V_{S_1}$ at $S_1$ is zero because the contact $S_1$ is the voltage source. Moreover,  
the intrinsic current fluctuation $\delta I_{S_1M}$ is zero due to vanishing temperatures of the modes between $S_1$ and $M$.
 Then, the correlation of the voltage fluctuation measured at $M$ reads
\begin{align}
 \overline{(\delta V_M )^2}= \left ( \frac{3h}{2e^2}\right )^2  \overline{(\delta I_{MS_{1/3}})^2}.
\end{align}
It proves Eq.~(13) in the main text. 
$\overline{(\delta I_{MS_{1/3}})^2}$ is given as Eq.~(1) of the main text with $L$ replaced by $L_{MS_{1/3}}$.
Note that the voltage fluctuation in the middle contact reflects the intrinsic noise generated in the line junction between $M$ and $S_{1/3}$.

\section{c. A line junction of couterpropagating modes $\delta \nu  = 1$ and $\delta \nu = -1$}
We consider the same setup as Fig.~\ref{LinejunctionMultiple} apart from that the filling factor $\nu$ is replaced by $1$.
Two counter-propagating modes are referred as $R$ and $L$: $R$ ($L$) stands for the right (left) moving modes.
The modes $R$ and $L$ are coupled to leads $S_{R}$ and $S_{L}$, respectively.
We take the same procedure as the previous sections:
we calculate the voltage and temperature profiles, and hence the zero frequency noise attaching artificial probe reservoirs to the edge modes between consecutive tunneling bridges.
Employing current conservation in reservoir $j$ ($1\leq j \leq N$), we set up equations
\begin{align} \label{currentconservation_supple}
 I_{R, j+1} (t) &=  I_{R ,j} (t)  + \left [ g \left (I_{L, j+1}(t)  -  I_{R, j} (t) \right ) +  I_{\tau, j}^{\rm{int}} (t) \right ] , \nonumber \\ 
I_{L, j+1} (t) &= I_{L, j} (t) +  \left [ g \left (I_{L, j+1}(t) -  I_{R, j} (t) \right ) +  I_{\tau, j}^{\rm{int}} (t) \right ] .  
\end{align}
From now, we focus on the case that a bias voltage is applied at $S_R$,
leaving $S_L$ grounded; $I_{R, 0}= e^2 V_0 / h$ and $I_{L, N+1} = 0$.
Taking the average of Eq.~\eqref{currentconservation_supple}, the voltage profile of each mode is obtained as
\begin{align} \label{voltageprofile}
 \overline{V_{R, j}} = V_0\frac{(N-j)g + 1}{Ng + 1},\,\,\,\,\,\,\,\,\,\,\,\,
 \overline{V_{L, j}} = V_0\frac{(N+1-j)g}{Ng + 1}.
\end{align}
Taking the continuum limit ($N \rightarrow \infty$) and defining the equilibration length $\ell_{\textrm{eq}}$ as $\ell_{\textrm{eq}}= (1- g) L / [(N+1)g]$, Eq.~\eqref{voltageprofile} becomes
\begin{align} \label{voltageprofilecontinuum}
V_{R} (x) = V_0\Big (\frac{\ell_{\textrm{eq}}+ L - x}{\ell_{\textrm{eq}}+ L}\Big ),\,\,\,\,\,\,\,\,\,\,\,\,
V_{L} (x) = V_0\Big ( \frac{L-x}{\ell_{\textrm{eq}}+ L}\Big ).
\end{align}
Then, the conductance from $S_{R}$ to $S_{L}$ is given as $G = e^2 V_{R}(L) / h = e^2 \ell_{\textrm{eq}} / [h ( \ell_{\textrm{eq}}  + L)]$.
Furthermore, the temperature profile is 
\begin{align}
k_B^2 T_{R}^2 (x) = \frac{3 (eV_0)^2}{\pi^2} \frac{x (\ell_{\textrm{eq}}+L - x)}{(\ell_{\textrm{eq}}+L)^2}, \,\,\,\,\,\,\,\,\,\,\,\,\,
k_B^2 T_{L}^2 (x) = \frac{3 (eV_0)^2}{\pi^2} \frac{(L-x) (\ell_{\textrm{eq}}+x)}{(\ell_{\textrm{eq}}+L)^2}.
\end{align}
Here, we have used the boundary conditions of $T_{R} (x = 0)= T_{L} (x = L) = 0$.

Under the condition that the voltage of the contacts is not fluctuating in time ($\delta I_{R, 0} (t) = \delta I_{L, N+1} (t) = 0$), the fluctuation $\delta I_{S_{L}} (t) = I_{S_{L}} (t) - \overline{I_{S_{L}}(t)}$ of 
the net current  at $S_{L}$ is written as
\begin{align}
\delta I_{S_{L}} (t) 
&= - \frac{1}{Ng +1 }  \sum_{j=0}^{N} \delta I_{\tau, j}^{\rm{int}} (t).
\end{align}
In the continuum limit ($N \rightarrow \infty$), the noise measured at $S_{L}$ is calculated as 
\begin{align}
\overline{(\delta I_{S_{L}})^2} = \frac{2e^2}{h} \frac{\ell_{\textrm{eq}}}{(\ell_{\textrm{eq}}+L)^2} \int_{0}^{L} dx [k_B T_R (x) + k_B T_L (x)] = 2 G eV_0 \frac{\sqrt{3}}{4}
=\frac{2e^2 (eV_0)}{h} \frac{\ell_{\textrm{eq}}}{( \ell_{\textrm{eq}}  + L)}\frac{\sqrt{3}}{4}.
\end{align}
This result coincides with the noise generated in a conventional metallic conductor with strong electron-electron interaction~\cite{Nagaev, Kozub, Dejong}. Note that this $1/L$ behavior of the noise differs from the $1/\sqrt{L}$ behavior of the noise in the line junction considered in the main text.  
The $1/L$ behavior of the noise is attributed to the fact that both electrial and thermal currents 
are transported in a diffusive way.

\section{d. five contact Setup}

\begin{figure} [h]
\includegraphics[width=0.9\columnwidth]{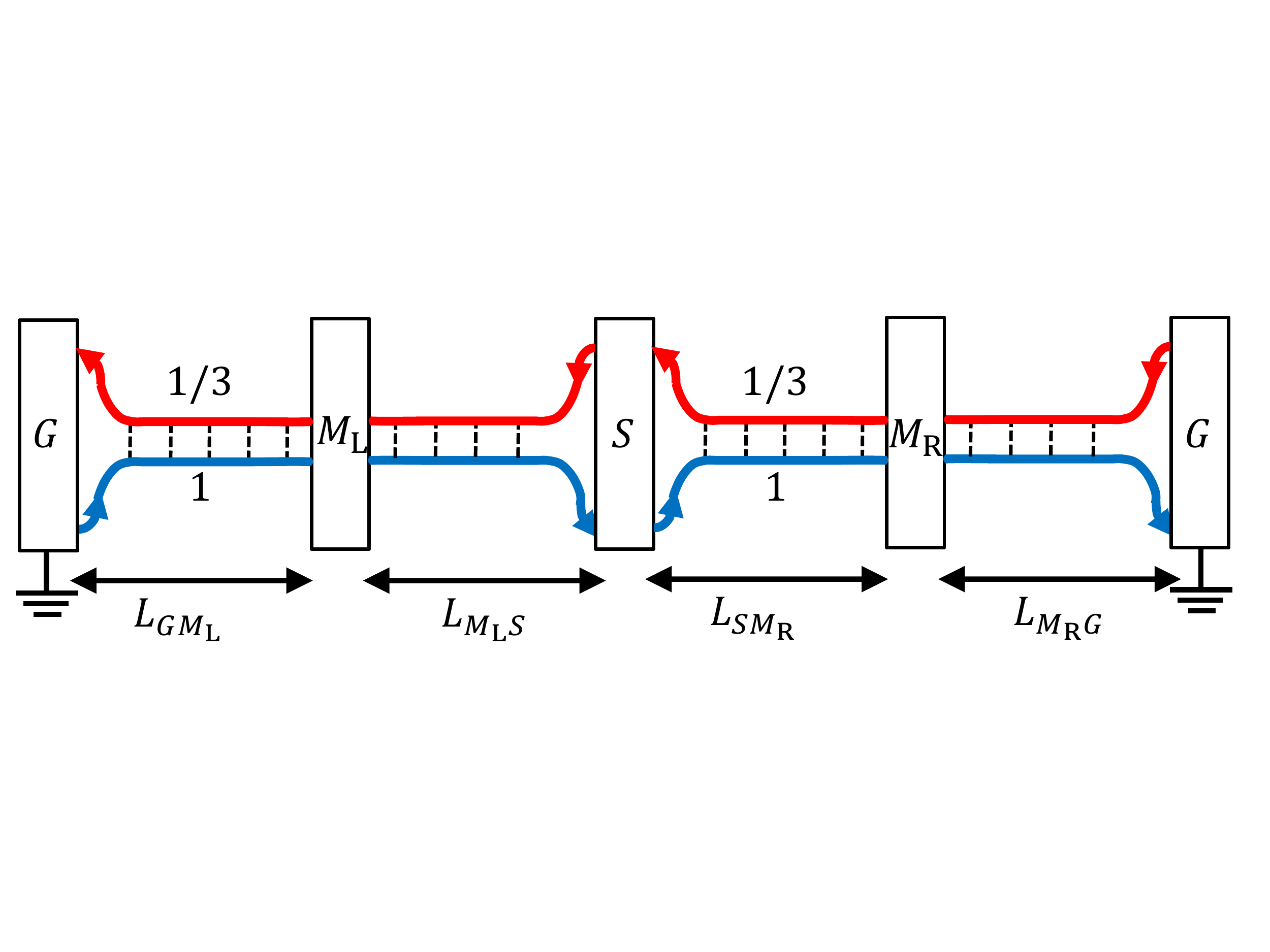} 
\caption{A setup with five contacts. A noiseless voltage (voltage source) is applied from $S$, and
the voltage fluctuation is measured in middle contacts $M_{\rm{L}}$ and $M_{\rm{R}}$. $M_{\rm{L}}$ and $M_{\rm{R}}$ are floating with respect to the a.c. component with the frequency on which the voltage fluctuation at $M$ is measured, as well as the d.c. component. Contacts denoted as $G$ are grounded.
} \label{5contactsSetup} 
\end{figure}

Finally, we address a setup with five contacts for experimental relevance (see Fig.~\ref{5contactsSetup}). Source $S$ is a noiseless voltage source and contacts $G$  are grounded. 
The voltage fluctuation is measured in middle contacts $M_{\rm{L}}$ and $M_{\rm{R}}$. $M_{\rm{L}}$ and $M_{\rm{R}}$ are floating with respect to the a.c. component of the voltage fluctuation with the frequency on which the voltage fluctuation at $M$ is measured, as well as the d.c. component. Since the voltage in the source is not fluctuating in time, the left side of $S$ is decoupled with 
the right side of $S$ in a sense of current fluctuation; I.e., the current fluctuation generated in the left side of $S$ is not transfered to the right side and vice versa. 
The right side of $S$ corresponds to the case that $S_{1}$ is biased while $S_{1/3}$ is grounded in the Fig.~1(b) of the main text. 
On the other hand, the left side of $S$ corresponds to the case that $S_{1/3}$ is biased while $S_{1}$ is grounded. 
Then, the zero frequency noise ($S_{M_{\rm{L}}}$ and $S_{M_{\rm{R}}}$) measured at $M_{\rm{L}}$ and $M_{\rm{R}}$
is written as 
\begin{align}
S_{M_{\rm{L}}}& \simeq \frac{c e^2}{h} (eV_0) \sqrt{\frac{\ell_{\textrm{eq}}}{L_{M_{\rm{L}} S}}},\,\,\,\,\,\,\,\,\,\,\,\,\,\,\,\,\,\,\,\,\,\,\,\,
S_{M_{\rm{R}}} \simeq \frac{c e^2}{h} (eV_0) \sqrt{\frac{\ell_{\textrm{eq}}}{L_{M_{\rm{R}} G}}}.
\end{align}
Note that the voltage fluctuation in the middle contacts reflects the intrinsic noise generated in the line junction, located at their just right side.

\begin{thebibliography}{10}
\bibitem{footnote} All the zero frequency correlation functions of current fluctuation are defined as $\overline{\delta I_{A} \delta I_B } \equiv 2 \int d\tau \delta I_{A} (t + \tau) \delta I_{B}(t) $, where $\delta I_{A(B)}(t) \equiv I_{A (B)} (t) - \overline{I_{A (B)}}$. Here the coefficient 2 in front of the integral is introduced for convention and
the time average of current $I_{A(B)}(t)$ is defined as 
$\overline{I_{A (B)}} = \lim_{\tau \rightarrow \infty}\int_{-\tau/2}^{\tau/2} dt I_{A(B)} (t) / \tau$.
\end{thebibliography}
\end{document}